\documentclass[review]{elsarticle}

\usepackage{lineno,hyperref}
\usepackage{mathtools}
\usepackage{amsmath}
\usepackage{color}
\usepackage{xcolor}
\usepackage{bm}
\usepackage{mathrsfs}
\usepackage{amssymb}
\usepackage{euscript}
\usepackage{algorithm}
\usepackage{algorithmic}
\usepackage{graphicx}
\usepackage{subfigure}
\usepackage{multirow}
\usepackage[Symbol]{upgreek}
\usepackage{booktabs}
\usepackage{makecell}
\usepackage[english]{babel}
\usepackage{diagbox}

\modulolinenumbers[1]

\journal{Applied Acoustics}

\begin{document}

\begin{frontmatter}

\title{Speech Enhancement using Progressive Learning-based Convolutional Recurrent Neural Network}
\author[mysecondaryddress,mythirdaddress]{Andong Li}
\author[myfourthaddress]{Minmin Yuan}
\author[mysecondaryddress,mythirdaddress]{Chengshi Zheng\corref{mycorrespondingauthor}}
\cortext[mycorrespondingauthor]{Corresponding author}
\ead{cszheng@mail.ioa.ac.cn}

\author[mysecondaryddress,mythirdaddress]{Xiaodong Li}

\address[mysecondaryddress]{Key Laboratory of Noise and Vibration Research, Institute of Acoustics, Chinese Academy of Sciences,
100190, Beijing, China}
\address[mythirdaddress]{University of Chinese Academy of Sciences, 100049, Beijing, China}
\address[myfourthaddress]{Research Institute of Highway Ministry of Transport, 100088, Beijing, China}




\begin{abstract}
Recently, progressive learning has shown its capacity to improve speech quality and speech intelligibility when it is combined with deep neural network (DNN) and long short-term memory (LSTM) based monaural speech enhancement algorithms, especially in low signal-to-noise ratio (SNR) conditions. Nevertheless, due to a large number of parameters and high computational complexity, it is hard to implement in current resource-limited micro-controllers and thus, it is essential to significantly reduce both the number of parameters and the computational load for practical applications. For this purpose, we propose a novel progressive learning framework with causal convolutional recurrent neural networks called PL-CRNN, which takes advantage of both convolutional neural networks and recurrent neural networks to drastically reduce the number of parameters and simultaneously improve speech quality and speech intelligibility. Numerous experiments verify the effectiveness of the proposed PL-CRNN model and indicate that it yields consistent better performance than the PL-DNN and PL-LSTM algorithms and also it gets results close even better than the CRNN in terms of objective measurements. Compared with PL-DNN, PL-LSTM, and CRNN, the proposed PL-CRNN algorithm can reduce the number of parameters up to 93\%, 97\%, and 92\%, respectively.
\end{abstract}

\begin{keyword}
speech enhancement\sep deep learning\sep progressive learning\sep convolutional neural network\sep long short-term memory
\end{keyword}

\end{frontmatter}


\section{Introduction}
Environmental noise is one of the major factors that have significant influences on robust automatic speech recognition (ASR), speech communication system and hearing implants~{\cite{loizou2013speech}}. To improve speech recognition accuracy and speech communication quality in real scenarios, speech enhancement algorithms are proposed to extract the desired speech from its noisy equivalent to improve signal-to-noise ratio (SNR). Conventional monaural speech enhancement methods include spectral subtraction~{\cite{boll1979suppression}}, Wiener filtering~{\cite{zheng13, chen2006new}}, subspace-based methods~{\cite{zheng14, jensen1995reduction}} and so on. It is well-known that the performance of these methods usually suffers from heavily decreased performance in extremely low SNR and non-stationary noise conditions, and it has already shown that these conventional speech enhancement algorithms cannot improve speech intelligibility for normal-hearing (NH) listeners~{\cite{healy2013algorithm}}.

In recent years, deep neural network (DNN)-based speech enhancement algorithms have attracted wide attention and their improved versions have been investigated owing to its superior potentials in complicated nonlinear mapping problems. DNN-based monaural speech enhancement algorithms are often categorized into two types, where one is the spectral mapping approach~{\cite{xu2014regression, fu2017complex}} and the other is the mask mapping approach~{\cite{wang2018supervised}}, which includes ideal binary mask (IBM)~{\cite{wang2005ideal}}, ideal ratio mask (IRM)~{\cite{hummersone2014ideal}}, ideal amplitude mask (IAM)~{\cite{wang2014training}} and complex domain-based mask~{\cite{williamson2016complex}}. In the previous fully connected (FC)-based noise reduction approaches, noisy features, and clean labels are extracted in the frame format so that speech is enhanced frame by frame. Because of neglecting the structural characteristics of speech spectra and long contexts relations between adjacent frames, these approaches often result in spectral artifacts and speech distortion in high-frequency bands. To resolve these problems, convolutional neural network (CNN)-based and recurrent neural network (RNN)-based models are recently investigated in the literature~{\cite{fu2017complex, park2017fully, fu2016snr, weninger2015speech, chen2017long}}, which have achieved better performance in both noise reduction and speech distortion. More recently, a type of advanced network named convolutional recurrent neural networks (CRNN) is proposed, which takes advantage of both CNNs and RNNs. Compared with the single-type network, previous experiments have shown that CRNN obtains better speech enhancement results~{\cite{zhao2018convolutional, tan2018convolutional}}.

Progressive learning (PL) has been introduced to the supervised speech enhancement algorithms~{\cite{gao2016snr, gao2018densely}}. Essentially, it can be regarded as a type of curriculum learning problem. Different from directly mapping from noisy features to clean targets, the whole process stage is divided into multiple easier and smaller stages, where the previous stages can boost the subsequent training processes. In~{\cite{gao2016snr}}, the whole stage is divided into multiple stages so that within each stage, its target is to improve SNR to some degrees instead of directly recovering clean speech. FC is utilized for training in each stage. In~{\cite{gao2018densely}}, LSTM is utilized as the principal layer, which can make full use of long-short time dependencies to obtain more spectral information. Note that, different from PL-DNN, dense connection is implemented in adjacent stages, which is helpful to improve the performance by effectively aggregating the information in different stages. For simplicity, architectures in the literature~{\cite{gao2016snr, gao2018densely}} are referred as PL-DNN and PL-LSTM, respectively.

This paper introduces a progressive learning model to improve the performance of CRNN, which can also significantly reduce its computational complexity and its number of parameters simultaneously. The proposed algorithm has several innovations when compared with previous works. First, causal CRNN is adopted as the sub-model within each stage instead of FC layers or LSTM layers. This is because CRNN is more parameter-efficient than LSTM~{\cite{tan2018convolutional}} and is capable for real-time applications when no future information is involved. Second, different from simply stacking independent and identical sub-net in each stage, the parameter sharing scheme is adopted, which can dramatically decrease the number of parameters while improving the performance in the regression task~{\cite{ren2019progressive}}. It is of vital importance for practical application and therefore, higher parameter efficiency can be achieved than previous networks. Finally, two training targets are chosen to study their influence on the performance of speech enhancement.

The remainder of this paper is organized as follows. In Section 2, the system flowchart is introduced briefly. Section 3 describes the training targets, and the CRNN-based model and its combination with progressive learning are also presented in this section. Section 4 presents the experimental settings. Section 5 gives the experimental results. The conclusion is presented in Section~6.

\section{PROBLEM FORMULATION AND SYSTEM OVERVIEW} \label{sec:system overview}
In the time domain, a mixture signal is typically modeled as
\begin{equation}
\setlength{\abovedisplayskip}{2pt}
\setlength{\belowdisplayskip}{2pt}
x\left( t \right) = s\left( t \right) + n\left( t \right),
\end{equation}
where $x\left( t \right)$, $s\left( t \right)$ and $n\left( t \right)$ refer to the noisy, the clean and the noise signals, respectively, in the time index $t$. From the perspective of frequency domain, the formula can be transformed into
\begin{equation}
\setlength{\abovedisplayskip}{2pt}
\setlength{\belowdisplayskip}{2pt}
X\left(k, l \right)= S\left(k, l \right) + N\left( k, l \right),
\end{equation}
where $X\left(k, l\right)$, $S\left(k, l\right)$ and $N\left(k, l\right)$ denote the noisy, the clean and the noise components, respectively, at the frequency bin index $k$ and the time frame index $l$. Monaural speech enhancement aims to extract the clean speech component from the observed noisy mixture. For spectral mapping-based approaches, the output is the estimated clean speech spectral magnitude, which is denoted as $\lvert\widehat S (k,l) \rvert$, where widehat $\widehat\bullet$ is used here to discriminate between the estimated and clean variables. For mask mapping-based approaches, mask $\widehat{M}(k,l)$ is estimated and can be used to estimate the clean speech spectral magnitude by multiplication operation.

\begin{figure}[t]
	\setlength{\abovecaptionskip}{0.235cm}
	\setlength{\belowcaptionskip}{-0.1cm}
	\centering
	\centerline{\includegraphics[width= \columnwidth]{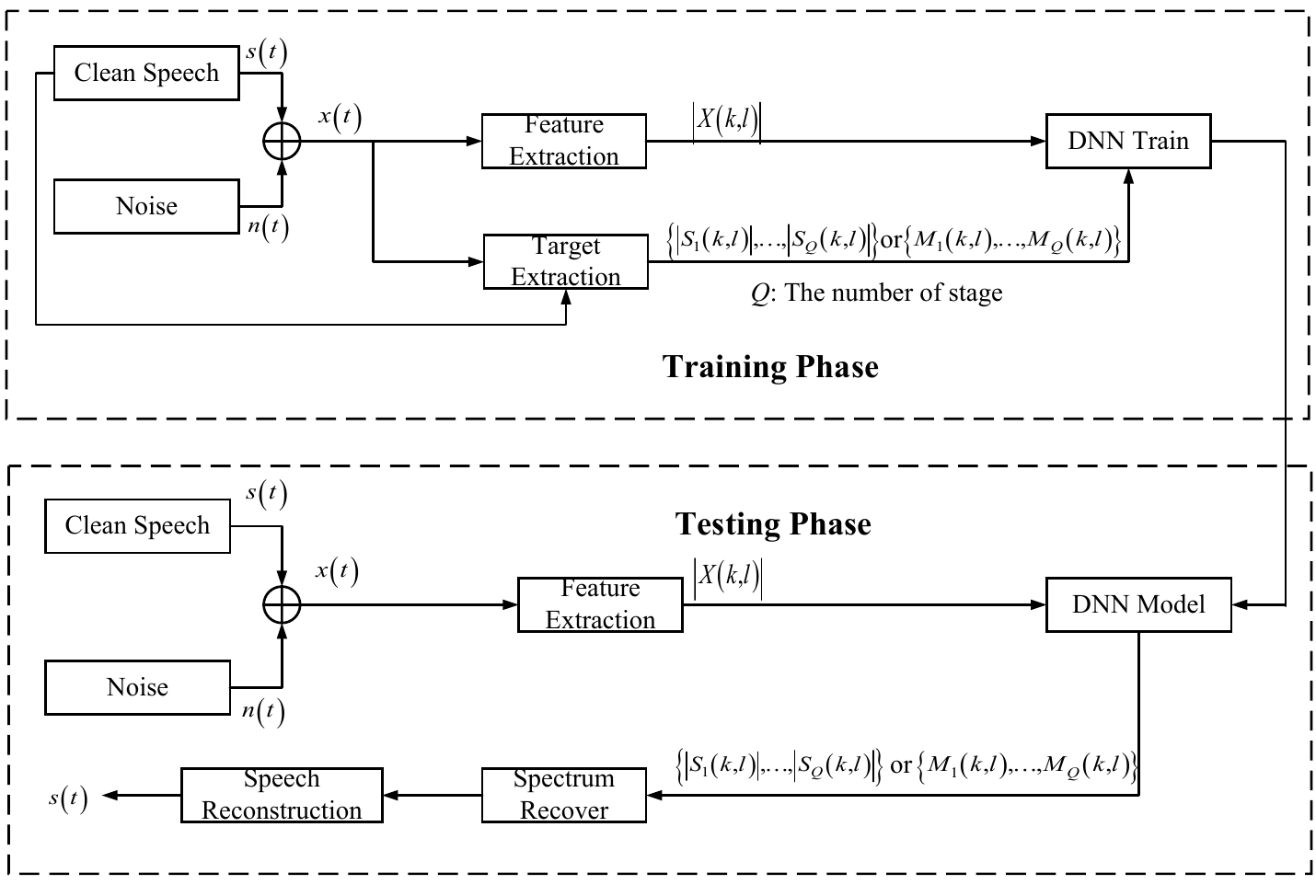}}
	\caption{ The block diagram of progressive learning.}
	\label{fig:schematic-flowchart}
	\vspace{0cm}
\end{figure}

The block diagram of progressive learning is shown in Figure~\ref{fig:schematic-flowchart}. In the training phase, clean speech signals and interference signals are mixed under various SNR conditions, where feature extraction and target calculation are followed. One can see that we use $\lvert X \left(k, l \right) \rvert$ to denote the feature and $\lvert S_{q} \left( k, l \right)\rvert$ with $q = 1 \dots Q$ to denote the target as an example. Here $Q$ is the number of stages. We adopt two types of targets herein, namely, the magnitude of the spectrum and the mask. The targets under different SNR-levels are sent to the deep learning model for training. In the testing phase, noisy features from the test dataset are fed into the well-trained deep learning model to estimate the desired targets in each T-F unit at different stages. After the speech spectral magnitude is estimated, the time-domain speech signal can be reconstructed with an overlap-add (OLA) technique.

\section{ARCHITECTURE}
\label{sec:learning-architecture}
\subsection{Features and Targets}
\begin{figure}[t]
	\centering
	\setlength{\abovecaptionskip}{0.235cm}
	\setlength{\belowcaptionskip}{-0.1cm}
	\centerline{\includegraphics[width= \columnwidth]{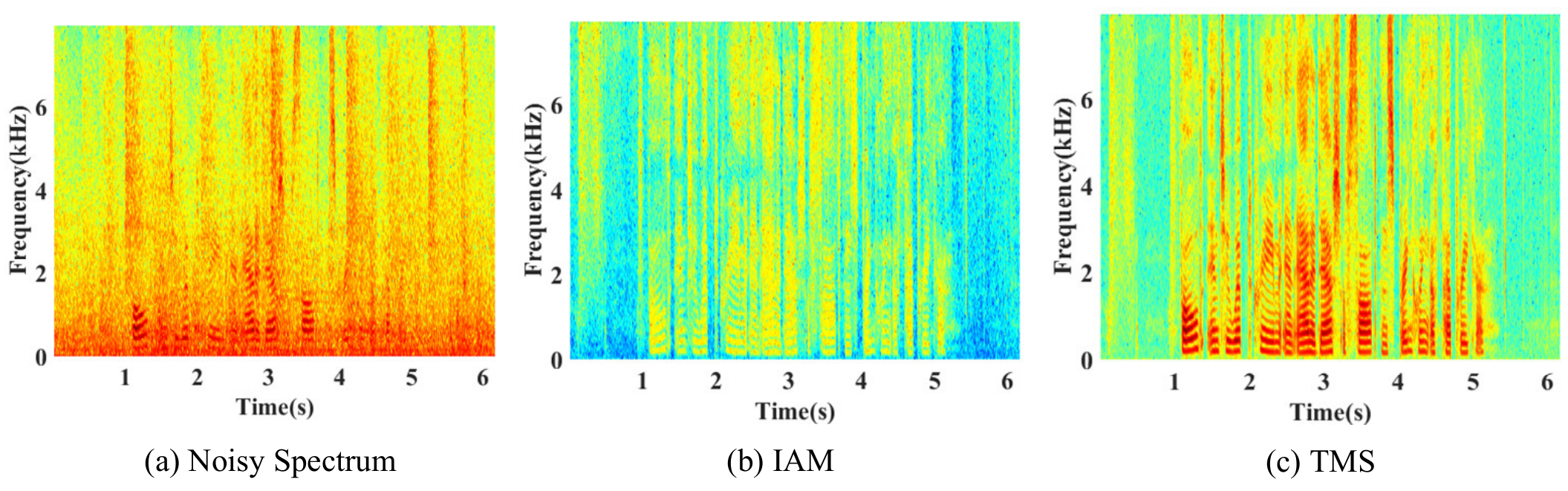}}
	\caption{Illustration of input and targets: (a) the spectrum of noisy speech, (b) ideal amplitude mask, (c) target magnitude spectrum. The training targets (b) and (c) are calculated under 0{\rm dB} condition, where the noisy speech is a mixture of a clean speech utterance taken from the TIMIT dataset and a factory noise taken from the NOISEX92 database.}
	\vspace{0cm}
	\label{fig:mask}
\end{figure}

In this study, two targets are considered, where one is the target magnitude spectrum (TMS) and the other is the ideal amplitude mask (IAM) with signal approximation (SA), which are illustrated in Figure~\ref{fig:mask}. Note that the noisy speech spectrum is also plotted as a reference. SA was introduced to directly optimize the source separation objective, which could reconstruct the speech signal in a best possible way~{\cite{weninger2014discriminatively}}. With SA, time-frequency (T-F) mask is estimated as the network output, which is subsequently multiplied with the corresponding noisy spectrum to obtain the estimated clean speech spectrum. While loss is then calculated between the estimated clean speech spectrum and the clean speech spectrum. Experiments have shown that SA often leads to better objective performance than direct mask optimization. To improve the training stability, we constrain the value of mask to range from 0 to 1 and the sigmoid function is chosen as the output nonlinearity accordingly. For TMS, as it is always non-negative, so the softplus function is adopted as the output activation function~{\cite{zheng2015improving}}. IAM with SA can be given by
\begin{equation}
\setlength{\abovedisplayskip}{2pt}
\setlength{\belowdisplayskip}{2pt}
\label{eqn:iam}
M\left( k, l\right) = \min \left\{\max \left\{ \left|\frac{S\left( k, l \right)} {X\left( k, l \right)}\right|, 0 \right\}, 1 \right\},
\end{equation}
\begin{equation}
\setlength{\abovedisplayskip}{2pt}
\setlength{\belowdisplayskip}{2pt}
\label{eqn:sa}
SA\left( k, l \right) = \left[S\left( k, l\right ) - X\left( k, l \right)\widehat{M}\left(k, l\right)\right]^2.
\end{equation}

\subsection{Causal Convolutional Recurrent Neural Network}
Causal CRNN is adopted as the sub-net in each stage. It resembles the architecture in {\cite{tan2018convolutional}} in which the principal part is the causal convolutional encoder-decoder (CED) with LSTM playing as a bottleneck layer to capture time dependencies. In the encoding part, the size of the feature map gradually decreases layer by layer, while in the decoding part the size gradually increases, indicating the compression and the extension of features.

Causal mechanism is applied for all the convolutional layers, which is first introduced in~{\cite{vanwavenet}} for generating raw audio waveform and it achieves state-of-the-art performance than all the previous models. During the calculation of the causal convolution, the predicted outputs do not use any future information and only the past temporal information is involved, and thus it is feasible for real-time processing applications.

RNNs with gated mechanism and long short-term memory have been proposed to mitigate gradient vanishing and exploding issues when learning long time sequences. LSTM is a type of typical memory unit, where different gates are set to control the percentage of saving, dropping temporal information and receiving incoming information.


\subsection{Proposed Architecture}
\begin{figure}[t]
	\setlength{\abovecaptionskip}{0.235cm}
	\setlength{\belowcaptionskip}{-0.1cm}
	\centering
	\centerline{\includegraphics[width= 100mm]{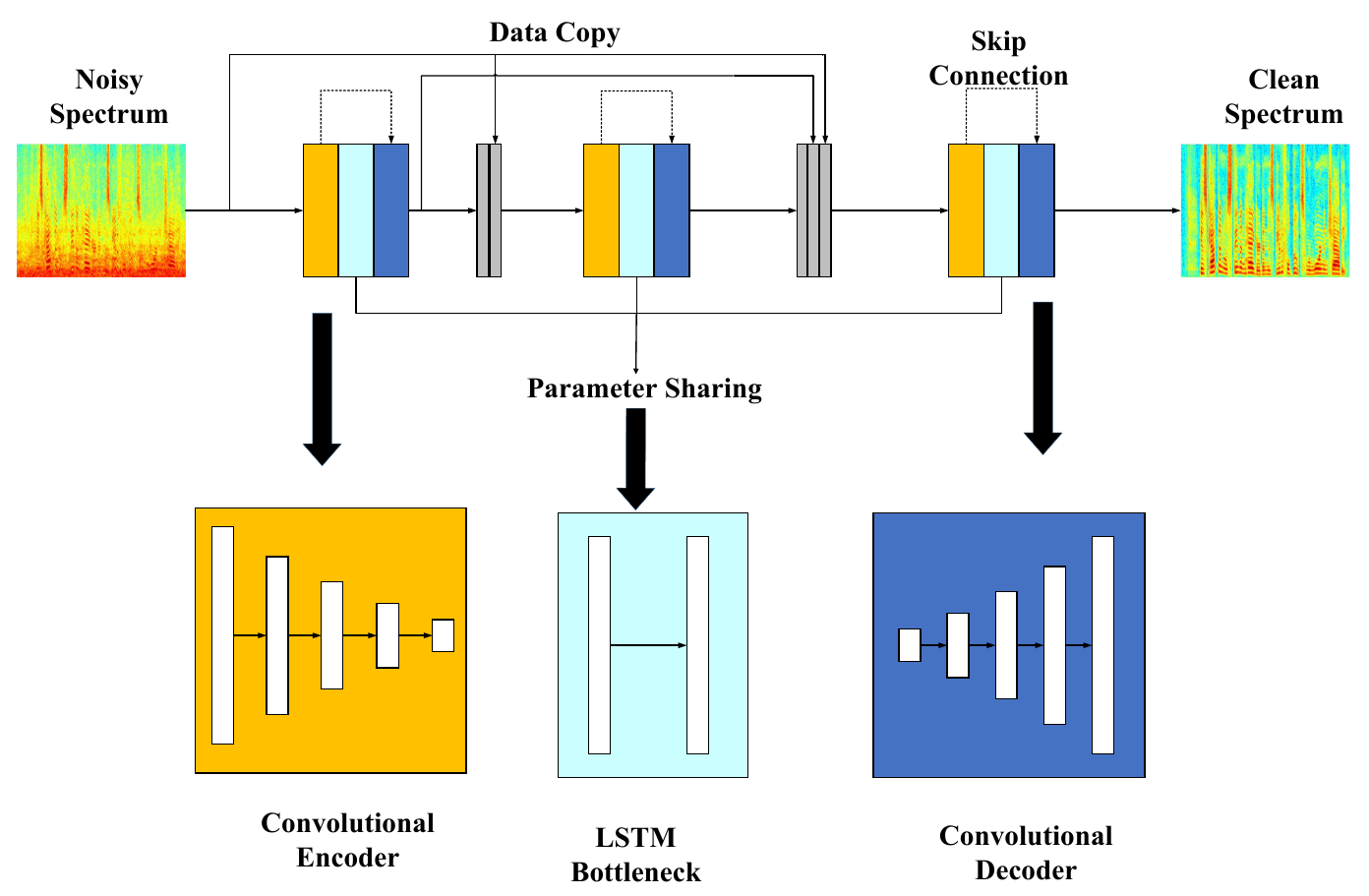}}
	\caption{The architecture of proposed progressive learning framework. Different modules are identified with different color blocks as shown at the bottom of the figure. Dashed line indicates the skip operation.}
	\vspace{0cm}
	\label{fig:architecture}
\end{figure}

Figure{~\ref{fig:architecture}} plots the architecture of PL-CRNN. The pipeline is comprised of multiple sub-nets, each of which is a type of causal CRNN but with much fewer parameters. The reason for utilizing fewer parameters within each stage is that PL is capable of compensating for the performance gap when a simplified network is adopted instead of the network with a larger number of parameters, which will be further analyzed in Section 5. The number of stages needs to be carefully chosen. This is because PL is expected to improve the performance, but too many learning stages may deteriorate the speech quality due to the information loss when a too deeper network is trained. Besides, more parameters and more computational complexity are inevitable with the increase of the number of stages. As shown in~{\cite{gao2018densely}}, $Q = 3$ could achieve a good trade-off between the performance and the number of parameters for PL-LSTM.


Within each stage, there are three main components, including the convolutional encoder, the LSTM bottleneck, and the convolutional decoder. For the encoding part, it consists of five convolutional blocks, which creates a compressed and deep representation of the input features by halving the size of feature dimension with striding operation in the frequency axes and doubling the number of channels layer. Each of the convolutional layer is followed by batch normalization (BN)~{\cite{ioffe2015batch}} and exponential linear unit (ELU)~{\cite{clevert2015fast}}. The decoder is the symmetric representation compared with the encoder, where the size of the frequency feature gradually increases by applying deconvolution~{\cite{noh2015learning}} and the number of channels decreases layer by layer. 	

As shown in ~\cite{kim2016deeply, ren2019progressive}, some parameters among different stages can be shared to reduce the network parameters while maintaining or improving its performance. This paper proposes to recursively utilize LSTM bottleneck in different stages to significantly reduce the number of parameters. While in both encoders and decoders, we do not share the parameters due to that their trainable parameters are much smaller than those of the LSTM module. Skip connection is introduced to compensate for the information loss during the feature compression process, which has been proved to be helpful for gradient flow. Additionally, different from the dense connection in~{\cite{gao2018densely}} that current outputs are concatenated with previous outputs in the feature axes, we choose to combine the outputs in the different stages along the channel axes, which could be beneficial for CNN training while effectively decreasing the additional parameters. Note that when IAM serves as the training target, before the current output is sent to the next stage, it needs to be multiplied with the noisy spectrum to obtain the estimated TMS.

A more detailed description of the proposed progressive learning architecture can refer to Table~\ref{tab1}. The whole network consists of three cascaded sub-models. As a result, only the sub-model parameter setting is detailed provided in the table. The input size and the output size of 3-D tensor representation are specified with $(Channels\times Timestep\times FrequencyFeat)$ format, while, for 2-D tensor representation, it is specified with $(Timestep\times Frequency Feat)$. The hyperparameters are specified with $Kernel Size$, $Stride$ and $Channel Number$ format. For stage $q$, the number of input channels is set to $q$ to satisfy the dense connection principle as shown in Table~\ref{tab1}.


{
	\renewcommand\arraystretch{0.7}
	\setlength{\tabcolsep}{5pt}
	\vspace{-0.3cm}
	\begin{table}[t]
		\tiny
		\footnotesize
		\caption{Detailed parameter setup of the proposed architecture. Our proposed model contains multiple stages for progressive noise reduction, where each stage has nearly the same structure except that the input channel of each stage is different. We only show the parameter setup and the tensor size of one stage for compact.}
		
		\begin{center}
			\begin{tabular}{c|c|c|c}
				\hline\hline
				layer name & input size & hyperparameters & output size\\
				\hline
				reshape\_size\_1 & $T$ $\times$ 161 & - & 1 $\times$ $T$ $\times$ 161\\
				\hline
				cascade\_1 & 1 $\times$ $T$ $\times$ 161 & - & $q$ $\times$ $T$ $\times$ 161\\
				\hline
				conv2d\_1 & $q$ $\times$ $T$ $\times$ 161 & 2 $\times$ 3, (1, 2), 16 & 16 $\times$ $T$ $\times$ 80\\
				\hline
				conv2d\_2 & 16 $\times$ $T$ $\times$ 80 & 2 $\times$ 3, (1, 2), 16 & 16 $\times$ $T$ $\times$ 39\\
				\hline
				conv2d\_3 & 16 $\times$ $T$ $\times$ 39 & 2 $\times$ 3, (1, 2), 16 & 16 $\times$ $T$ $\times$ 19\\
				\hline
				conv2d\_4 & 16 $\times$ $T$ $\times$ 19 & 2 $\times$ 3, (1, 2), 32 & 32 $\times$ $T$ $\times$ 9\\
				\hline
				conv2d\_5 & 32 $\times$ $T$ $\times$ 9 & 2 $\times$ 3, (1, 2), 64 & 64 $\times$ $T$ $\times$ 4\\
				\hline
				reshape\_size\_2& 64 $\times$ $T$ $\times$ 4 & - & T $\times$ 256\\
				\hline
				lstm1 (reuse) &$T$ $\times$ 256 & 256 & $T$ $\times$ 256\\
				\hline
				lstm2 (reuse) &$T$ $\times$ 256 & 256 & $T$ $\times$ 256\\
				\hline
				reshape\_size\_3 & $T$ $\times$ 256 & - & 64 $\times$ $T$ $\times$ 4\\
				\hline
				deconv2d\_1 & 128 $\times$ $T$ $\times$ 4 & 2 $\times$ 3, (1, 2), 32 & 32 $\times$ $T$ $\times$ 9\\
				\hline
				deconv2d\_2 & 64 $\times$ $T$ $\times$ 9 & 2 $\times$ 3, (1, 2), 16 & 16 $\times$ $T$ $\times$ 19\\
				\hline
				deconv2d\_3 & 32 $\times$ $T$ $\times$ 19 & 2 $\times$ 3, (1, 2), 16 & 16 $\times$ $T$ $\times$ 39\\
				\hline
				deconv2d\_4 & 32 $\times$ $T$ $\times$ 39 & 2 $\times$ 3,(1, 2), 16 & 16 $\times$ $T$ $\times$ 80\\
				\hline
				deconv2d\_5 & 32 $\times$ $T$ $\times$ 80 & 2 $\times$ 3, (1, 2), 1 & 1 $\times$ $T$ $\times$ 161\\
				\hline
				reshape\_size\_4 &1 $\times$ $T$ $\times$ 161 & - & $T$ $\times$ 161\\
				\hline\hline
			\end{tabular}
			\label{tab1}
		\end{center}
	\end{table}
}

\section{EXPERIMENTS}
\subsection{Datasets Preparation}
Extensive experiments are conducted with the TIMIT corpus, which include 630 speakers of eight major dialects of American English with each reading tens utterances. 2000, 400 and 100 utterances are selected as the training, evaluation and testing datasets, respectively. Training dataset are mixed under different SNR levels ranging from -5\rm{dB} to 10\rm{dB} with the interval 1\rm{dB} while testing datasets are mixed under (-5\rm{dB}, 0\rm{dB}, 5\rm{dB}, 10\rm{dB}). 115 types of noises~{\cite{xu2014regression, hu2010tandem}} are utilized for training. To explore the generalization capacity of the model, five types of unseen test noises from NOISEX92~{\cite{varga1993assessment}} are utilized, including babble, f16, factory2, m109 and white.

Various noises are first concatenated into a long vector. During the mixing process, the random cutting point is selected to obtain the noise-only signal, which is subsequently mixed with a clean utterance under one random chosen SNR condition. As a result, a total of 20,000, 2,000 and 800 contaminated utterances and their clean version pairs are built for training, evaluation and testing, respectively. Table~{\ref{snr-impro}} summarizes multiple intermediate targets used in the following experiments with different SNR improvement levels over the originally noisy utterance.

{
	\renewcommand\arraystretch{0.7}
	\setlength{\tabcolsep}{5pt}
	\vspace{-0cm}
	\begin{table}[t]
		\tiny
		\footnotesize
		\caption{SNR improvement setting of intermediate stages for different $Q$}
		
		\begin{center}
			\begin{tabular}{c|c}
				\hline\hline
				Stage number $Q$ & intermediate SNR improvement\\
				\hline
				2 & +20{\rm dB}\\
				\hline
				3 & +10{\rm dB}, +20{\rm dB}\\
				\hline
				4 & +5{\rm dB}, +10{\rm dB}, +20{\rm dB}\\
				\hline
				5 & +5{\rm dB}, +10{\rm dB}, +15{\rm dB}, +20{\rm dB}\\
				\hline\hline
			\end{tabular}
			\label{snr-impro}
		\end{center}
	\end{table}
}

The feature extraction approach takes as follows. For the sampling rate 16 kHz, the 20-ms Hamming window function is applied with 10-ms overlap. A 320-point STFT is adopted, leading to a 161-D feature vector for each frame. Instead of using the logarithm power spectrum (LPS), we use the magnitude as the feature directly, which is the same as~{\cite{tan2018gated}} for fair comparison. Besides, we use the same values of $\alpha_q$ as~{\cite{gao2016snr}}, where $\alpha_{q} = 0.1$ for $q= 1,\cdots, Q-1$ and $\alpha_{Q} = 1$.

\subsection{Comparison Models}
In our experiments, we compare our proposed model with another four baselines, i.e., progressive learning with feedforward DNN~{\cite{gao2016snr}}, LSTM-based progressive learning with dense connection~{\cite{gao2018densely}}, autoencoder-based CRNN in real-time {\cite{tan2018convolutional}} and its simplified version with smaller training parameters, which are termed as PL-DNN, PL-LSTM, CRNN, and SCRNN, respectively. Follow the literature~{\cite{gao2018densely}}, $Q = 3$ is chosen to compare previous PL-based networks with the proposed PL-CRNN. In addition, PL-CRNN with $Q = 5$ is also presented to compare with the latter two baselines, i.e., CRNN and SCRNN. All the models are configured as the causal models, given as follows:

\begin{enumerate}[(1)]
\item PL-DNN. Slightly different from the literature~{\cite{gao2016snr}}, we take the causal strategy for a fair comparison and the previous 10 frames are combined together with the current frame to form a larger feature vector, i.e., $161\times11 = 1771$. As a consequence, the structure of PL-DNN is $\{2048-2048-161-2048-2048-161-2048-2048-161\}$.

\item PL-LSTM. The structure of PL-LSTM is $\{1024-1024-161-1024-1024-161-1024-1024-161\}$, where 1024 refers to LSTM layer and 161 is the affine layer. In addition, dense connection is utilized in adjacent stages and post-processing (PP) is applied to both PL-DNN and PL-LSTM by averaging the outputs in different stages, which have been proved to further improve its performance~{\cite{gao2018densely}}.

\item CRNN. It has a similar structure with the proposed sub-net but with much more trainable parameters, which is known as one of the state-of-the-art models. The output channels of the encoder are 256 and the number of LSTM units is 1024.

\item SCRNN. Compared with CRNN, it decreases the number of parameters for both CNN and LSTM modules. That is to say, it has about a quarter of the trainable parameters compared with CRNN. SCRNN is presented here to show the performance degradation when reducing the number of parameters of CRNN in a simple way.
\end{enumerate}

\subsection{Optimization Details}
All the models are trained with Adam optimizer~{\cite{kingma2014adam}}. The learning rate is initialized at 0.001. We halve the learning rate only when consecutive 3 evaluation loss increment arises and the training process is early-stopped only if more than 10 increment on evaluation loss happens. We train the model for 100 epochs to guarantee the model convergence. The minibatch is set to 16 at an utterance level. Within a minibatch, zero value is padded for all the utterances whose timestep length is less than the longest one. For PL-DNN, utterances from a minibatch are reshaped into $(Batch\times Timestep\times FrequencyFeat)$ format to meet the required input size.

\section{RESULTS AND ANALYSIS}
\subsection{Evaluation Metrics}
In this study, we evaluate the performance of different models in terms of speech quality and speech intelligibility with three objective metrics, containing perceptual evaluation of speech quality (PESQ)~{\cite{rix2001perceptual}}, short-time objective intelligibility (STOI)~{\cite{taal2011algorithm}} and source-to-distortion ratio (SDR)~{\cite{vincent2006performance}}.

\subsection{the Influence of Multi-stage}
\begin{figure}[t]
	\setlength{\abovecaptionskip}{0.235cm}
	\setlength{\belowcaptionskip}{-0.1cm}
	\centering
	\subfigure[]{
		\includegraphics[width= 55mm]{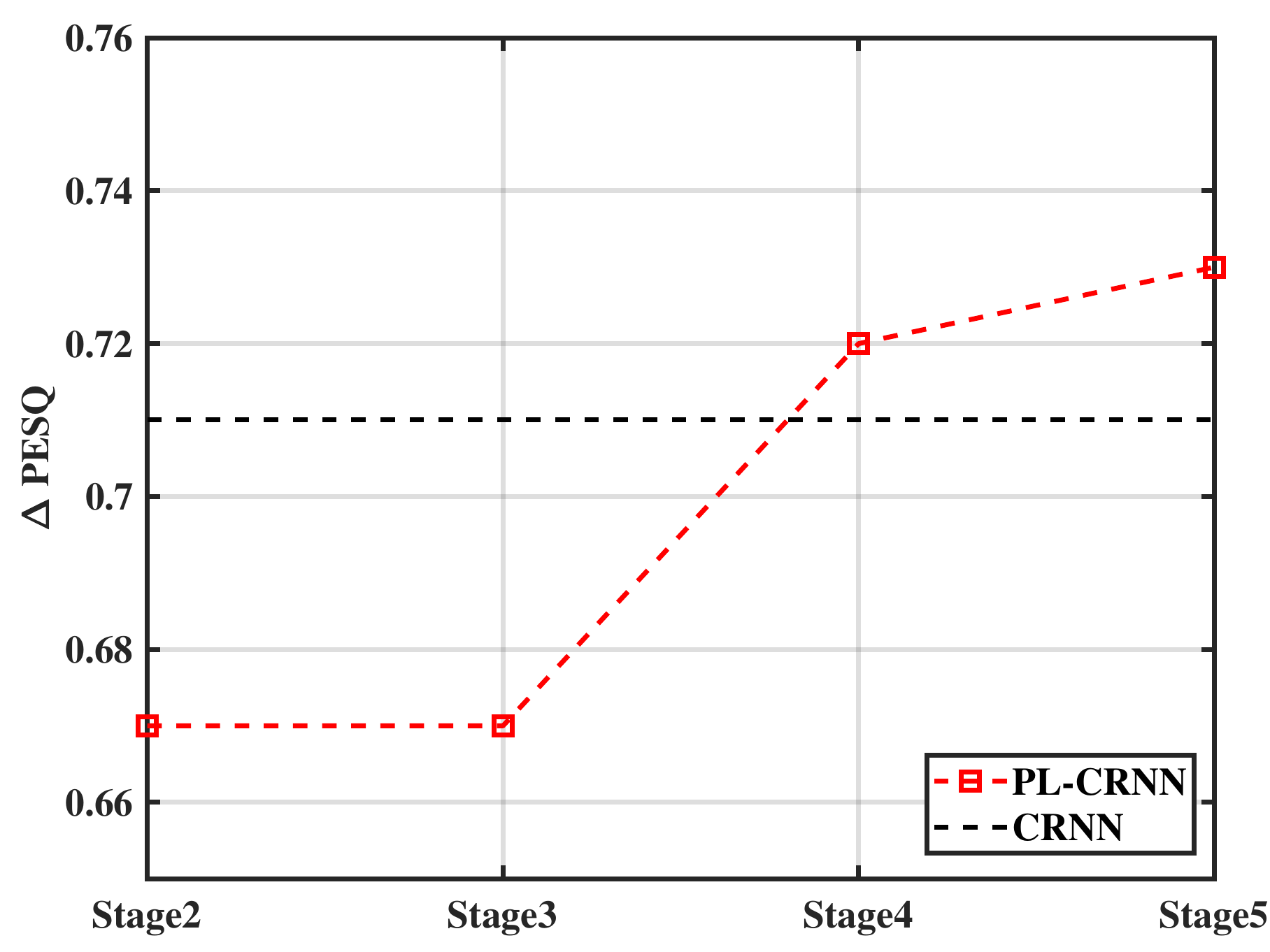}	
	}
	\subfigure[]{
		\includegraphics[width= 55mm]{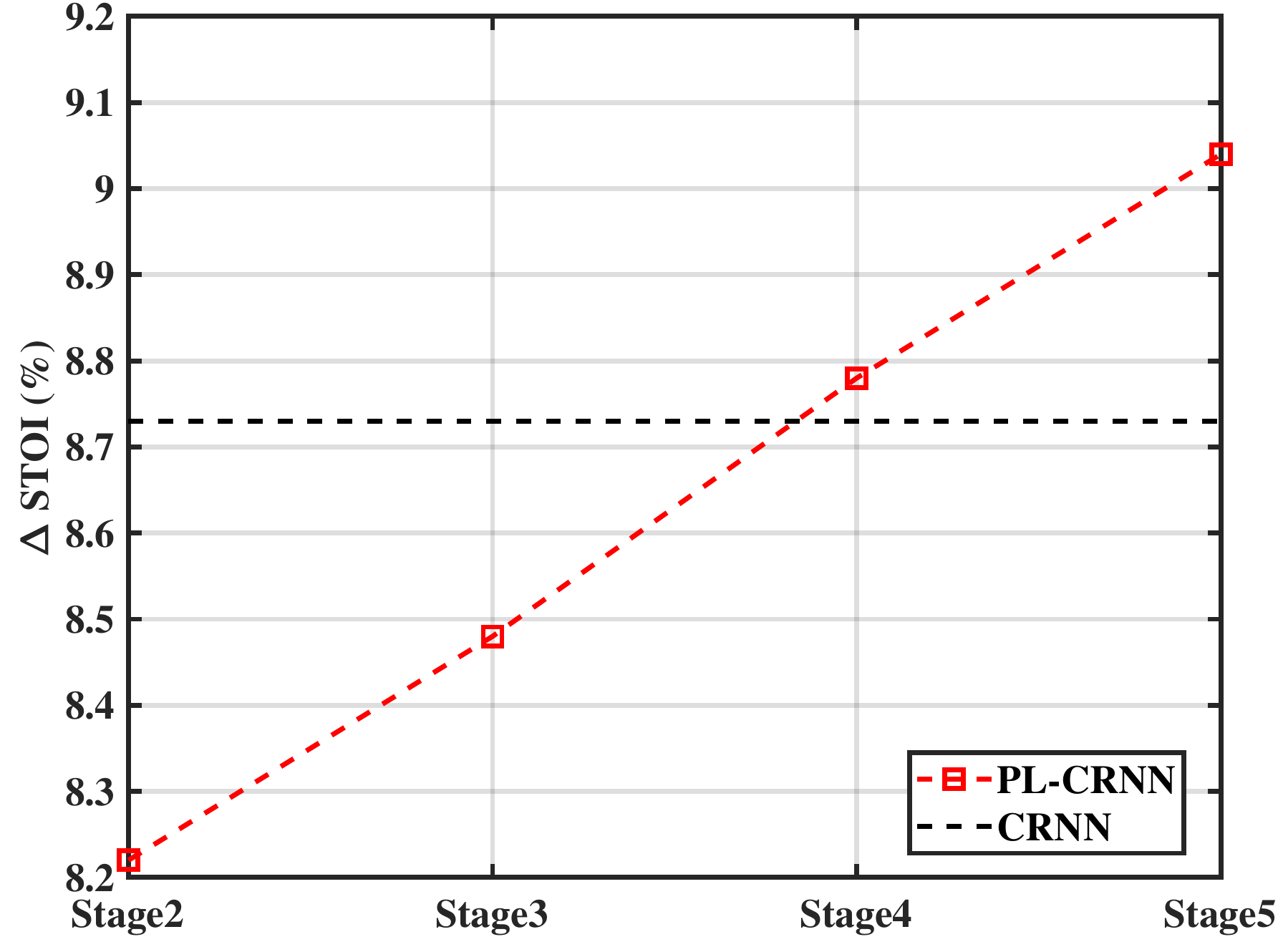}	
	}
	\caption{Average metric improvement in terms of PESQ and STOI for PL-CRNN and CRNN. TMS is trained as the target. (a) $\Delta$PESQ versus different numbers of stages. (b) $\Delta$STOI versus different numbers of stages.}
	\label{fig:stage-improvement}
	\vspace{0cm}
\end{figure}

We first demonstrate the influence of multi-stage from the perspective of the number of stages $Q$. As stated before, the previous stages can boost performance in the subsequent task. Generally, when the number of stages increases, it is expected to improve the performance. We train the network PL-CRNN with a smaller training dataset under different number of stages to study its influence on speech enhancement. The training set consists of 10,000 noisy-clean pairs that are randomly selected from the original training dataset. TMS is the training target. The testing results are averaged under both seen and unseen noise conditions, which are presented in Figure~{\ref{fig:stage-improvement}}. The results of CRNN are also given for comparison.
	
From Figure.~{\ref{fig:stage-improvement}}, one can observe that both metrics achieve an overall improvement with the increase of the number of stages. Specifically, when $Q \ge 4$, PL-CRNN has better performance than CRNN in terms of both PESQ and STOI scores. Additionally, as parameter-efficient encoder and decoder are utilized for each stage, the increase of the number of stages will introduce only very limited extra parameters, which will be studied in the last part of Section 5. However, for PL-DNN and PL-LSTM, the increase of the number of stages will introduce inefficient FC and LSTM layers.
	
\subsection{Comparison of Objective Results}
Table~{\ref{tbl:seen-results1}} and Table~{\ref{tbl:unseen-results1}} summarize the results of different models for seen and unseen cases, respectively. Two targets (TMS and IAM) are studied for both baselines and the proposed architecture. $Q = 3$ is chosen if not specially noted for PL-based models.
\subsubsection{Comparison with previous PL-based models}
We first compare the results among PL-based models with the same number of stages. From the tables, we can have the following remarks. (1) Both PL-DNN and PL-LSTM can effectively improve speech quality and intelligibility for both seen and unseen noise, which is consistent with the results in~{\cite{gao2016snr, gao2018densely}}. (2) Compared with PL-DNN, PL-LSTM achieves relatively better performance in terms of all the three metrics, which can be explained from three perspectives. One is that the LSTM layer is capable of utilizing sequence dependencies from previous timesteps while the DNN layer only captures contextual information with a preset frame window, which heavily limits the performance. Another is that dense operation along the frequency axes is helpful to aggregate the information from the previous stages. The other is that ReLU is helpful for gradient flow when exploring a depth network compared with the sigmoid function. (3) A notable improvement in three metrics is observed for PL-CRNN compared with the baselines. For example, when going from PL-LSTM to PL-CRNN, an average PESQ improvement of 0.18 is observed for unseen case while about 0.12 improvement for unseen case. The reasons behind the results can be explained as higher efficiency can be achieved when the sub-net in each stage is CRNN instead of simple FC or LSTM layers. On one hand, convolutional auto-encoder is applied, which facilitates the spectral patterns to be better explored and learned during the training. On the other hand, recursive LSTM layers are applied as the sequence modeling part, which can effectively model the sequence correlation in different stages while dramatically decreasing the trainable parameters. Besides, the usage of dense connection along channel axes introduces negligibly extra parameters compared with dense connection along feature axes.

\renewcommand\arraystretch{0.8}
\newcolumntype{V}{!{\vrule width 2pt}}
\begin{table*}[t]
	\caption{Experimental results with seen noise under different SNR conditions. \textbf{BOLD} indicates the best result for each case. Average values are also given for better comparison. TMS and IAM are two targets in this study. The number of stages is $Q = 3$ for all PL-based algorithm if no emphasis is given.}
	\centering
	\LARGE
	\resizebox{\textwidth}{!}{
		
		\begin{tabular}{cVcVcccccVcccccVccccc}
			\Xhline{2pt}
			\multicolumn{2}{cV}{\textbf{Metrics}} & \multicolumn{5}{cV}{\textbf{PESQ}}  & \multicolumn{5}{cV}{\textbf{STOI (in \%)}}
			& \multicolumn{5}{c}{\textbf{SDR (in dB)}}  \\
			\Xhline{2pt}
			\multicolumn{2}{cV}{\textbf{SNR (dB)}} & \textbf{-5}  &\textbf{0} &\textbf{5} &\textbf{10} &\textbf{Avg.}
			& \textbf{-5}  &\textbf{0} &\textbf{5} &\textbf{10} &\textbf{Avg.}
			& \textbf{-5}  &\textbf{0} &\textbf{5} &\textbf{10} &\textbf{Avg.}
			\\
			\Xhline{2pt}
			\multicolumn{2}{cV}{\textbf{Noisy}}  &1.43 &1.77 &2.16 &2.48 &1.96 &62.24 &72.22 &81.29 &88.82 &76.15 &-4.59 &0.20 &5.10 &10.07 &2.69\\
			\Xhline{2pt}
			\multirow{6}*{\textbf{TMS}}
			&  \textbf{PL-DNN} &1.88 &2.38 &2.70 &2.98 &2.49 &74.54 &83.46 &89.02 &92.75 &84.94 &5.47 &9.10 &12.61 &15.15 &10.58\\
			&  \textbf{PL-LSTM} &2.08 &2.47 &2.79 &3.08 &2.60 &76.25 &85.07 &90.44 &94.09 &86.46 &5.90 &9.52 &13.11 &16.24 &11.19\\
			&  \textbf{PL-CRNN} &2.28 &2.64 &2.96 &3.21 &2.77 &78.57 &86.80 &91.61 &94.92 &87.98 &7.77 &10.75 &14.29 &17.39 &12.55\\
			&  \textbf{CRNN} &2.29 &2.67 &2.98 &3.25 &2.80 &78.47 &86.71 &91.68 &94.89 &87.94 &7.71 &10.76 &14.29 &17.42 &12.54\\
			& \textbf{SCRNN} &2.20 &2.59 &2.91 &3.18 &2.72 &77.25 &85.97 &90.88 &94.21 &87.08 &7.22 &9.97 &13.65 &16.80 &11.91 \\
			&  \textbf{PL-CRNN (Q = 5)} &$\mathbf{2.36}$ &$\mathbf{2.71}$ &$\mathbf{3.00}$ &$\mathbf{3.25}$ &$\mathbf{2.83}$ &$\mathbf{79.60}$ &$\mathbf{87.43}$ &$\mathbf{92.08}$ &$\mathbf{95.21}$ &$\mathbf{88.58}$ &$\mathbf{8.14}$ &$\mathbf{11.24}$ &$\mathbf{14.50}$ &$\mathbf{17.63}$ &$\mathbf{12.88}$\\
			\Xhline{2pt}
			\multirow{6}*{\textbf{IAM}}
			&	\textbf{PL-DNN} &2.00 &2.38 &2.70 &2.99 &2.52 &74.30 &83.17 &89.12 &93.29 &84.97 &5.39 &9.19 &12.96 &16.34 &10.97\\
			&	\textbf{PL-LSTM} &2.09 &2.46 &2.79 &3.07 &2.60 &75.62 &84.60 &90.14 &94.15 &86.13 &5.81 &9.52 &13.35 &16.83 &11.38\\
			&	\textbf{PL-CRNN} &2.31 &2.67 &2.98 &3.24 &2.80 &78.50 &86.63 &91.49 &94.74 &87.84 &7.84 &10.83 &14.38 &17.47 &12.63 \\
			&	\textbf{CRNN} &2.32 &2.67 &2.99 &3.26 &2.81 &78.71 &86.64 &91.46 &94.81 &87.90 &7.88 &10.84 &14.39 &17.48 &12.65\\
			&	\textbf{SCRNN} &2.25 &2.61 &2.93 &3.20 &2.75 &77.12 &85.64 &90.57 &94.06 &86.85 &7.32 &10.31 &13.87 &17.01 &12.13\\
			&   \textbf{PL-CRNN (Q = 5)} &$\mathbf{2.38}$ &$\mathbf{2.72}$ &$\mathbf{3.01}$ &$\mathbf{3.26}$ &$\mathbf{2.84}$ &$\mathbf{79.35}$ &$\mathbf{87.21}$ &$\mathbf{91.69}$ &$\mathbf{94.99}$ &$\mathbf{88.31}$ &$\mathbf{8.00}$ &$\mathbf{11.28}$ &$\mathbf{14.58}$ &$\mathbf{17.65}$ &$\mathbf{12.88}$\\
			\Xhline{2pt}
	\end{tabular}}
	\label{tbl:seen-results1}
	\vspace{0.1cm}
\end{table*}

\renewcommand\arraystretch{0.8}
\newcolumntype{V}{!{\vrule width 2pt}}
\begin{table*}[t]
	\caption{Experimental results with unseen noise under different SNR conditions. \textbf{BOLD} indicates the best result for each case. Average values are also given for better comparison. TMS and IAM are two targets in this study. The number of stages is $Q = 3$ for all PL-based models if no emphasis is given.}
	\centering
	\LARGE
	\resizebox{\textwidth}{!}{
		
		\begin{tabular}{cVcVcccccVcccccVccccc}
			\Xhline{2pt}
			\multicolumn{2}{cV}{\textbf{Metrics}} & \multicolumn{5}{cV}{\textbf{PESQ}}  & \multicolumn{5}{cV}{\textbf{STOI (in \%)}}
			& \multicolumn{5}{c}{\textbf{SDR (in dB)}}  \\
			\Xhline{2pt}
			\multicolumn{2}{cV}{\textbf{SNR (dB)}} & \textbf{-5}  &\textbf{0} &\textbf{5} &\textbf{10} &\textbf{Avg.}
			& \textbf{-5}  &\textbf{0} &\textbf{5} &\textbf{10} &\textbf{Avg.}
			& \textbf{-5}  &\textbf{0} &\textbf{5} &\textbf{10} &\textbf{Avg.}
			\\
			\Xhline{2pt}
			\multicolumn{2}{cV}{\textbf{Noisy}} &1.49 &1.80 &2.19 &2.49 &1.99 &59.67 &71.22 &81.45 &89.24 &75.40 &-4.80 &0.10 &5.07 &10.06 &2.61\\
			\Xhline{2pt}
			\multirow{6}*{\textbf{TMS}}
			&  \textbf{PL-DNN}  &1.84 &2.26 &2.63 &2.90 &2.41 &68.63 &79.98 &87.41 &92.00 &82.00 &3.10 &7.80 &11.50 &14.82 &9.30\\
			&  \textbf{PL-LSTM} &1.96 &2.38 &2.74 &3.00 &2.52 &71.48 &82.33 &89.32 &93.20 &84.08 &3.92 &8.51 &12.20 &15.79 &10.11\\
			&  \textbf{PL-CRNN} &2.06 &2.50 &2.85 &3.13 &2.63 &73.16 &83.42 &90.15 &94.14 &85.32 &5.38 &9.59 &13.01 &16.77 &11.18\\
			&  \textbf{CRNN} &2.07 &2.51 &2.87 &3.15 &2.65 &72.51 &83.42 &90.09 &94.09 &85.03 &5.10 &9.44 &12.98 &16.78 &11.07\\
			&  \textbf{SCRNN} &2.02 &2.45 &2.82 &3.08 &2.59 &72.24 &82.94 &89.36 &93.38 &84.48 &4.86 &9.06 &12.48 &16.22 &10.65\\
			&  \textbf{PL-CRNN (Q = 5)} &$\mathbf{2.13}$ &$\mathbf{2.55}$ &$\mathbf{2.90}$ &$\mathbf{3.16}$ &$\mathbf{2.69}$ &$\mathbf{73.58}$ &$\mathbf{84.10}$ &$\mathbf{90.52}$ &$\mathbf{94.44}$ &$\mathbf{85.66}$ &$\mathbf{5.62}$ &$\mathbf{9.83}$ &$\mathbf{13.30}$ &$\mathbf{16.99}$ &$\mathbf{11.43}$\\
			\Xhline{2pt}
			\multirow{6}*{\textbf{IAM}}
			&	\textbf{PL-DNN} &1.81 &2.23 &2.62 &2.89 &2.39 &67.69 &79.36 &87.29 &92.38 &81.68 &3.00 &7.76 &11.78 &15.79 &9.58\\
			&	\textbf{PL-LSTM} &1.94 &2.35 &2.72 &2.99 &2.50 &70.20 &81.19 &88.68 &93.34 &83.35 &3.64 &8.25 &12.18 &16.12 &10.05\\
			&	\textbf{PL-CRNN} &2.05 &2.51 &2.87 &3.15 &2.64 &72.15 &83.33 &89.71 &93.88 &84.77 &5.28 &9.55 &13.05 &16.82 &11.18\\
			&	\textbf{CRNN} &2.07 &2.51 &2.88 &3.16 &2.65 &72.80 &83.27 &89.76 &93.88 &84.93 &5.25 &9.51 &13.00 &16.76 &11.13\\
			&	\textbf{SCRNN} &2.02 &2.45 &2.81 &3.06 &2.59 &71.45 &82.36 &88.90 &93.25 &83.99 &4.81 &8.98 &12.45 &16.25 &10.62\\
			&	\textbf{PL-CRNN (Q = 5)} &$\mathbf{2.13}$ &$\mathbf{2.56}$ &$\mathbf{2.90}$ &$\mathbf{3.17}$ &$\mathbf{2.69}$ &$\mathbf{72.59}$ &$\mathbf{83.54}$ &$\mathbf{89.89}$ &$\mathbf{94.03}$ &$\mathbf{85.02}$ &$\mathbf{5.68}$ &$\mathbf{9.85}$ &$\mathbf{13.23}$ &$\mathbf{16.95}$ &$\mathbf{11.43}$\\
			\Xhline{2pt}
	\end{tabular}}
	\label{tbl:unseen-results1}
	\vspace{0.1cm}
\end{table*}

\begin{figure}[t]
	\setlength{\abovecaptionskip}{0.235cm}
	\setlength{\belowcaptionskip}{-0.1cm}
	\centering
	\subfigure[]{
		\includegraphics[width= 55mm]{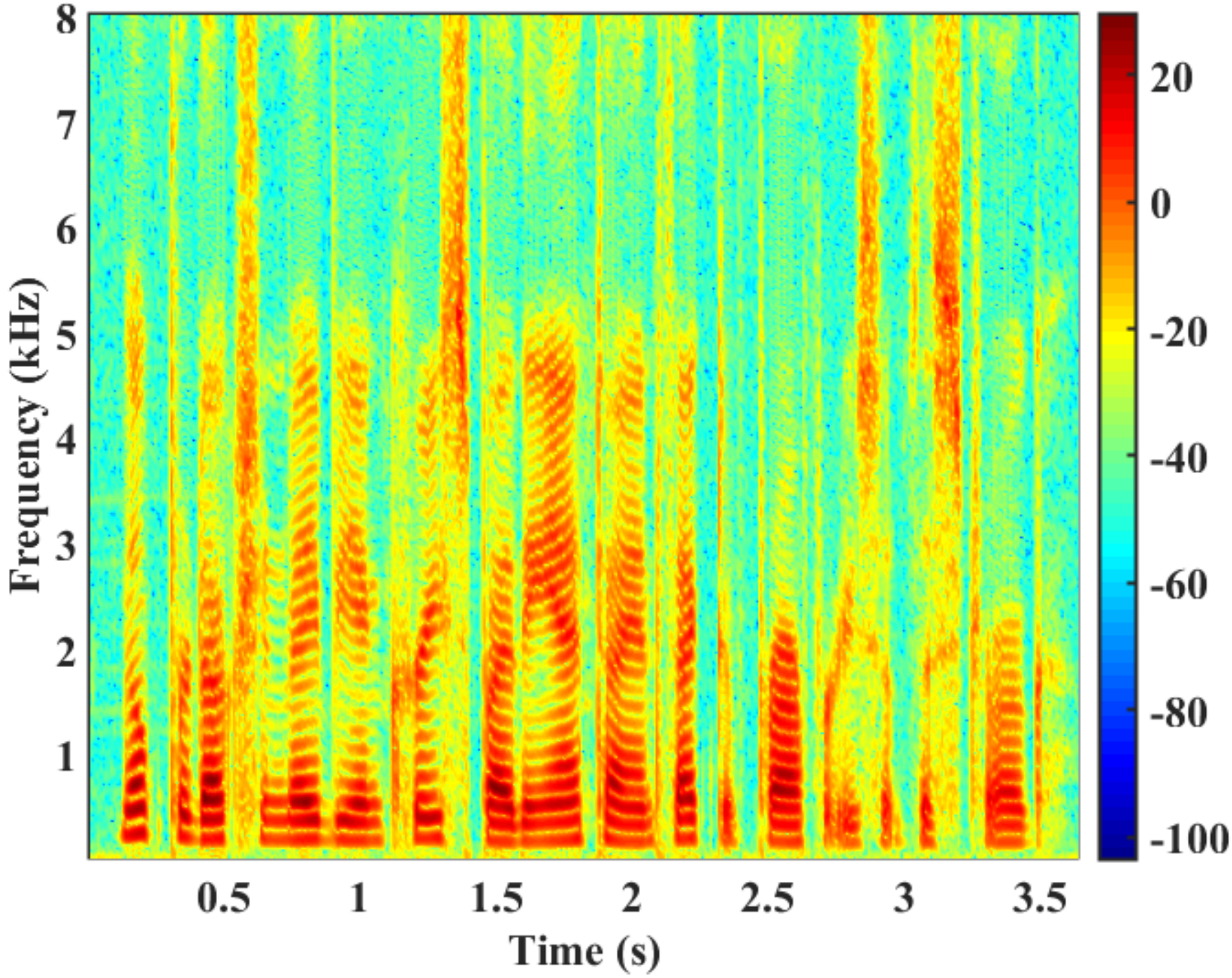}	
	}
	\subfigure[]{
		\includegraphics[width= 55mm]{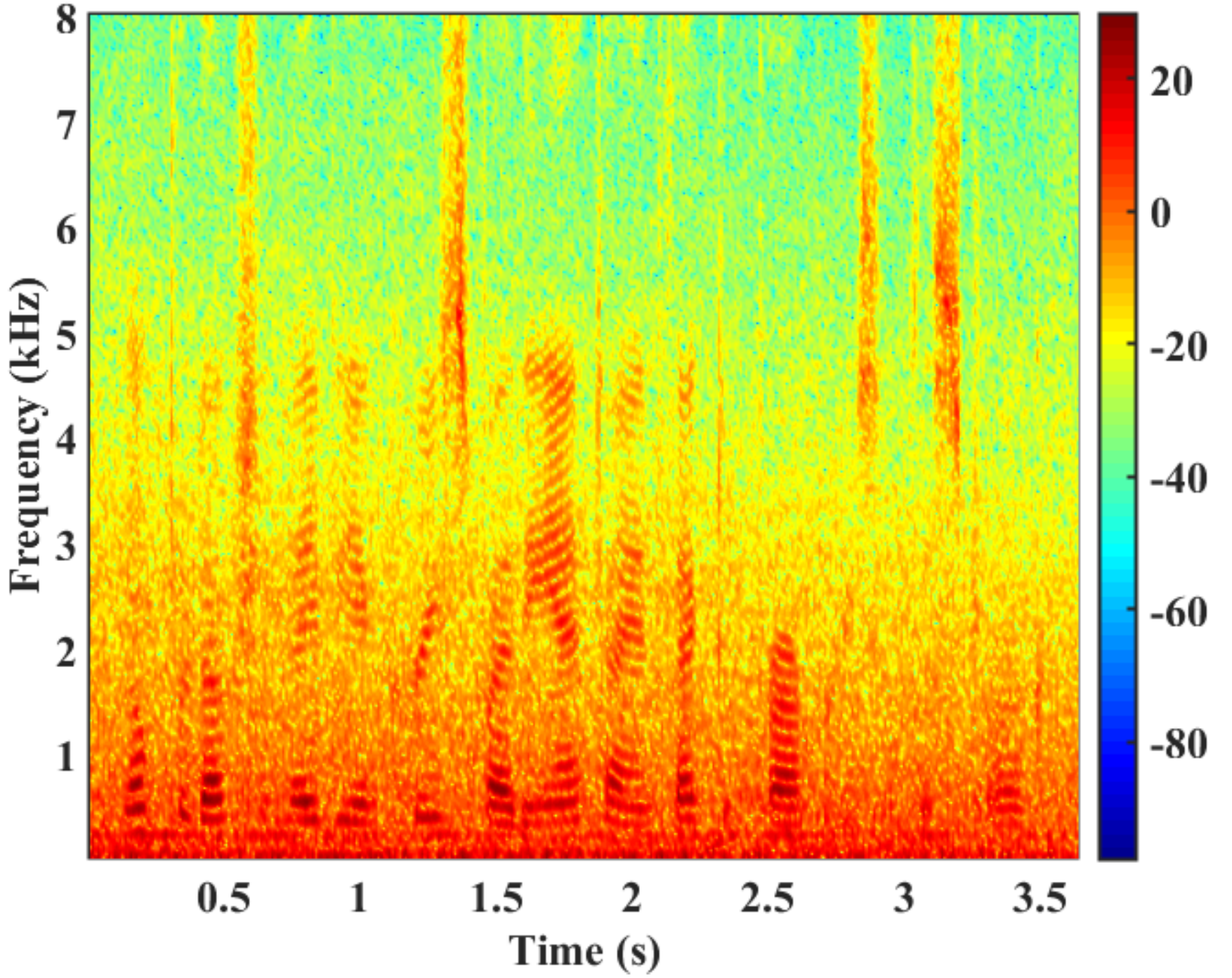}	
	}
	\subfigure[]{
		\includegraphics[width= 55mm]{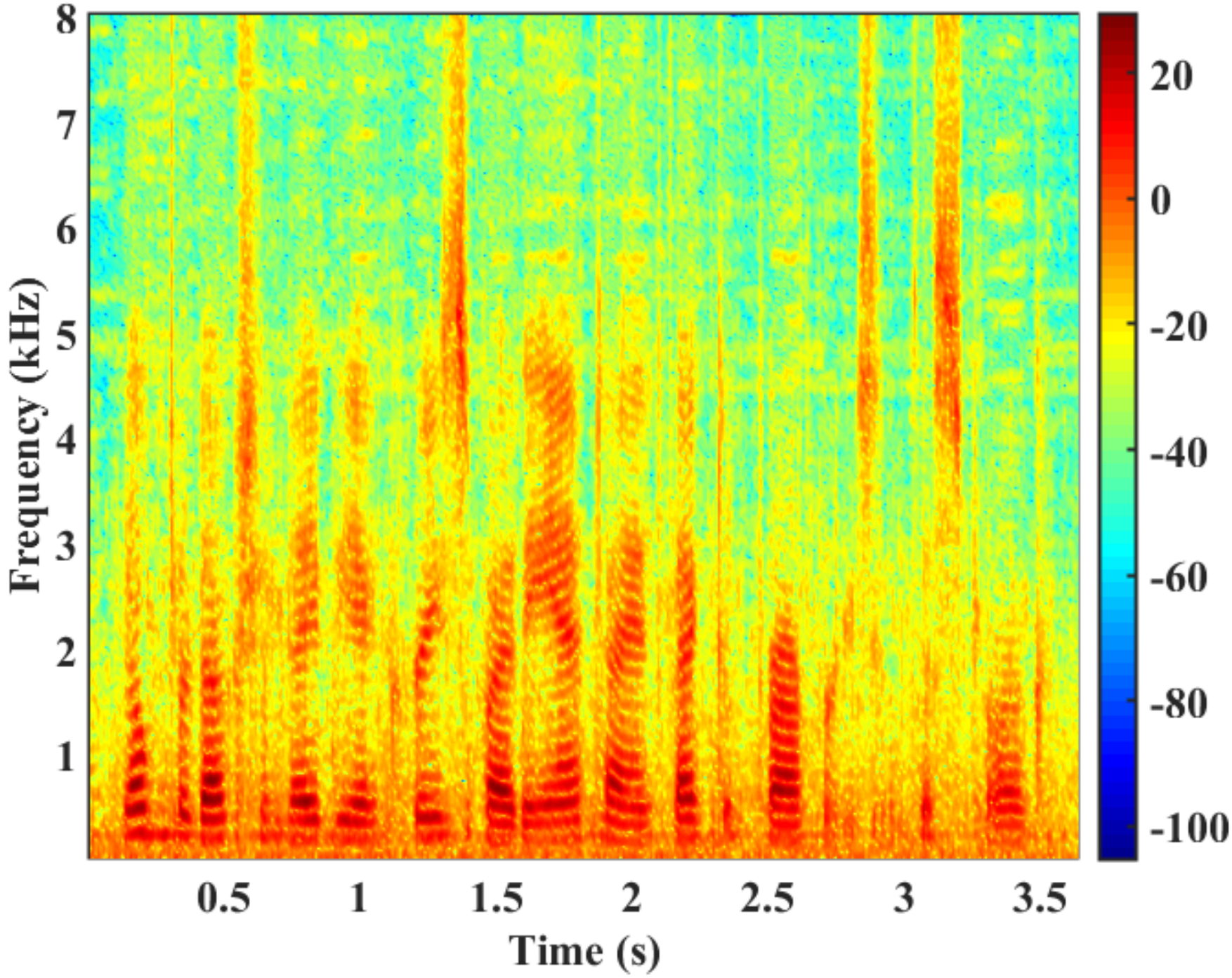}	
	}
	\subfigure[]{
		\includegraphics[width= 55mm]{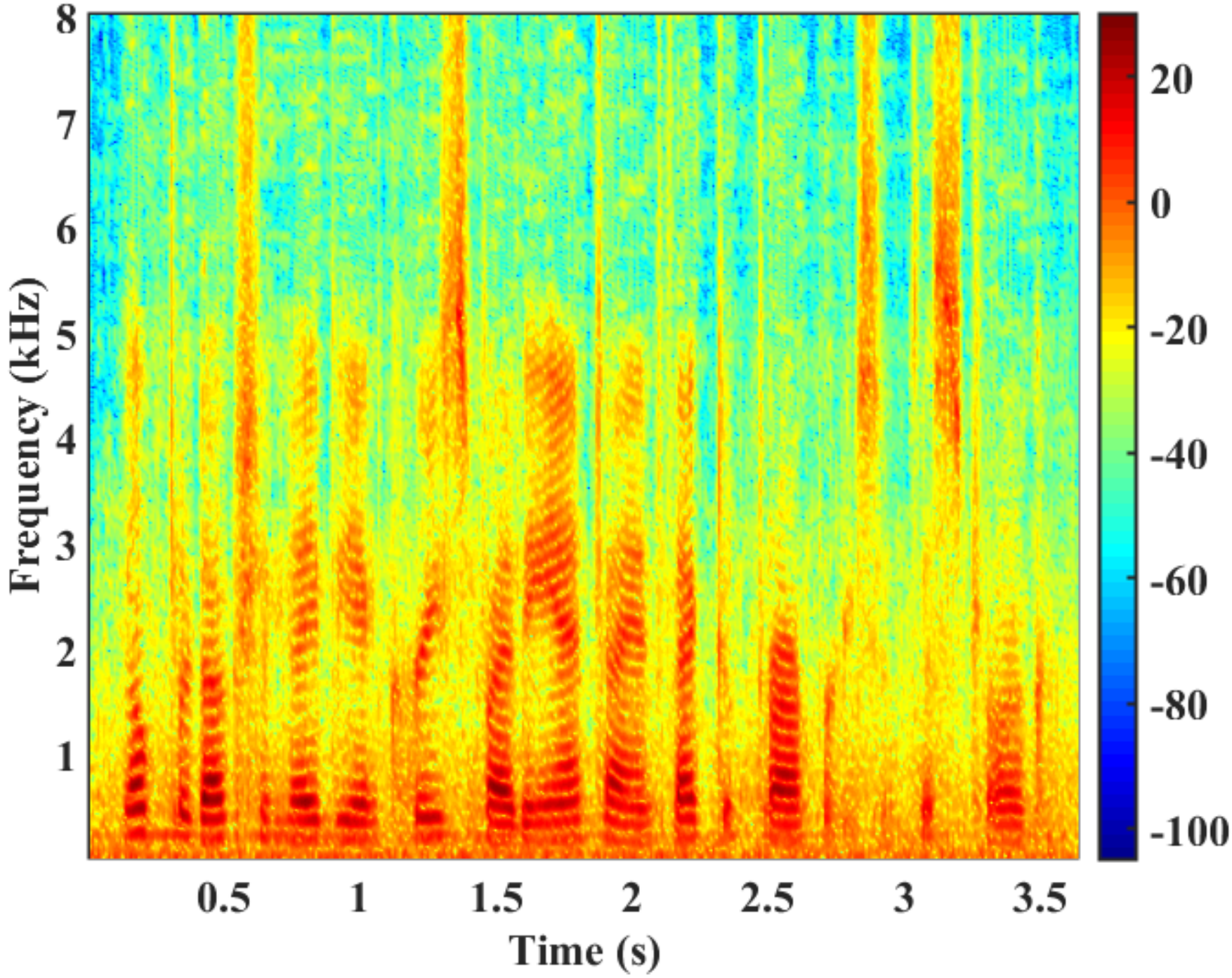}	
	}
	\subfigure[]{
		\includegraphics[width= 55mm]{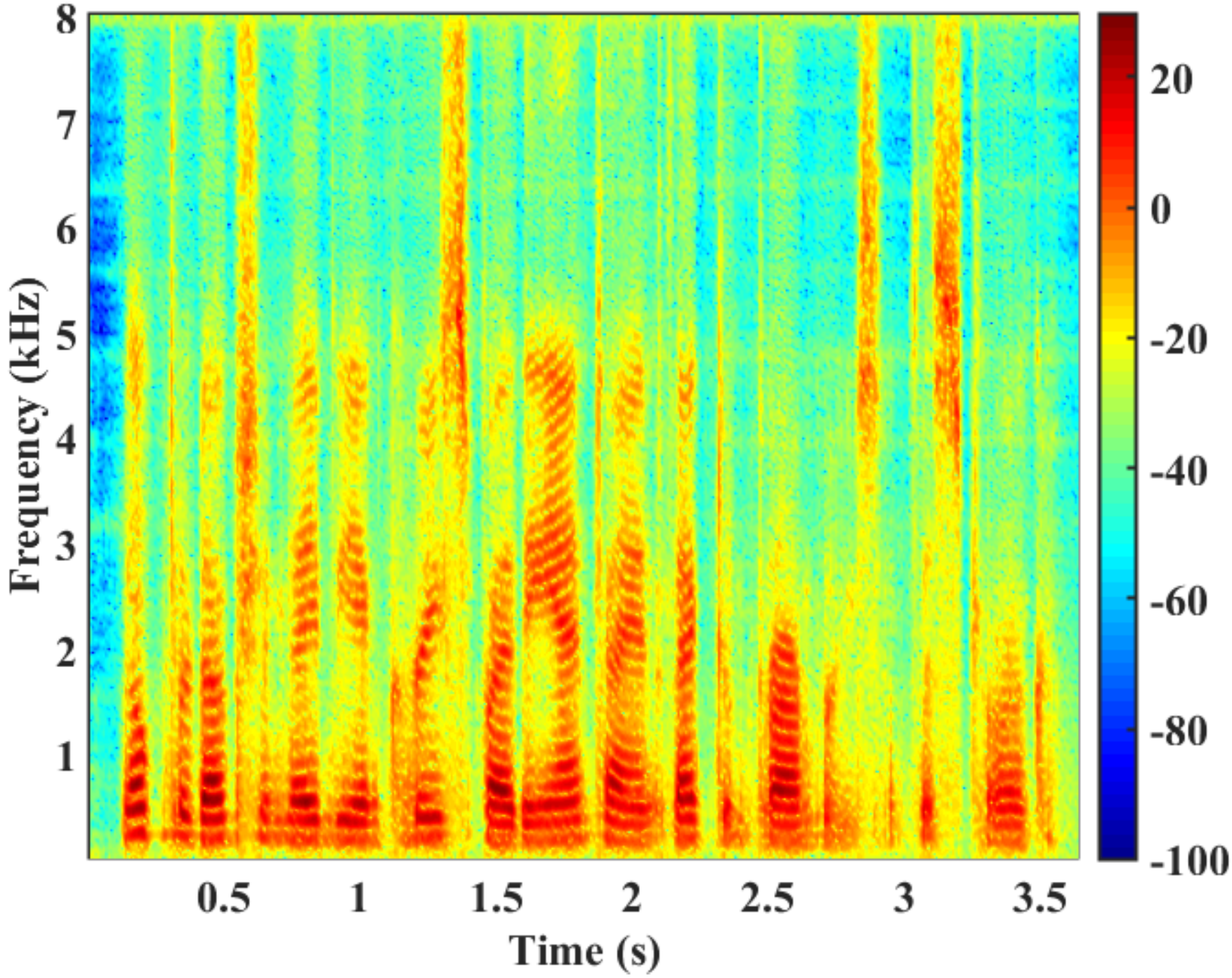}	
	}
	\subfigure[]{
		\includegraphics[width= 55mm]{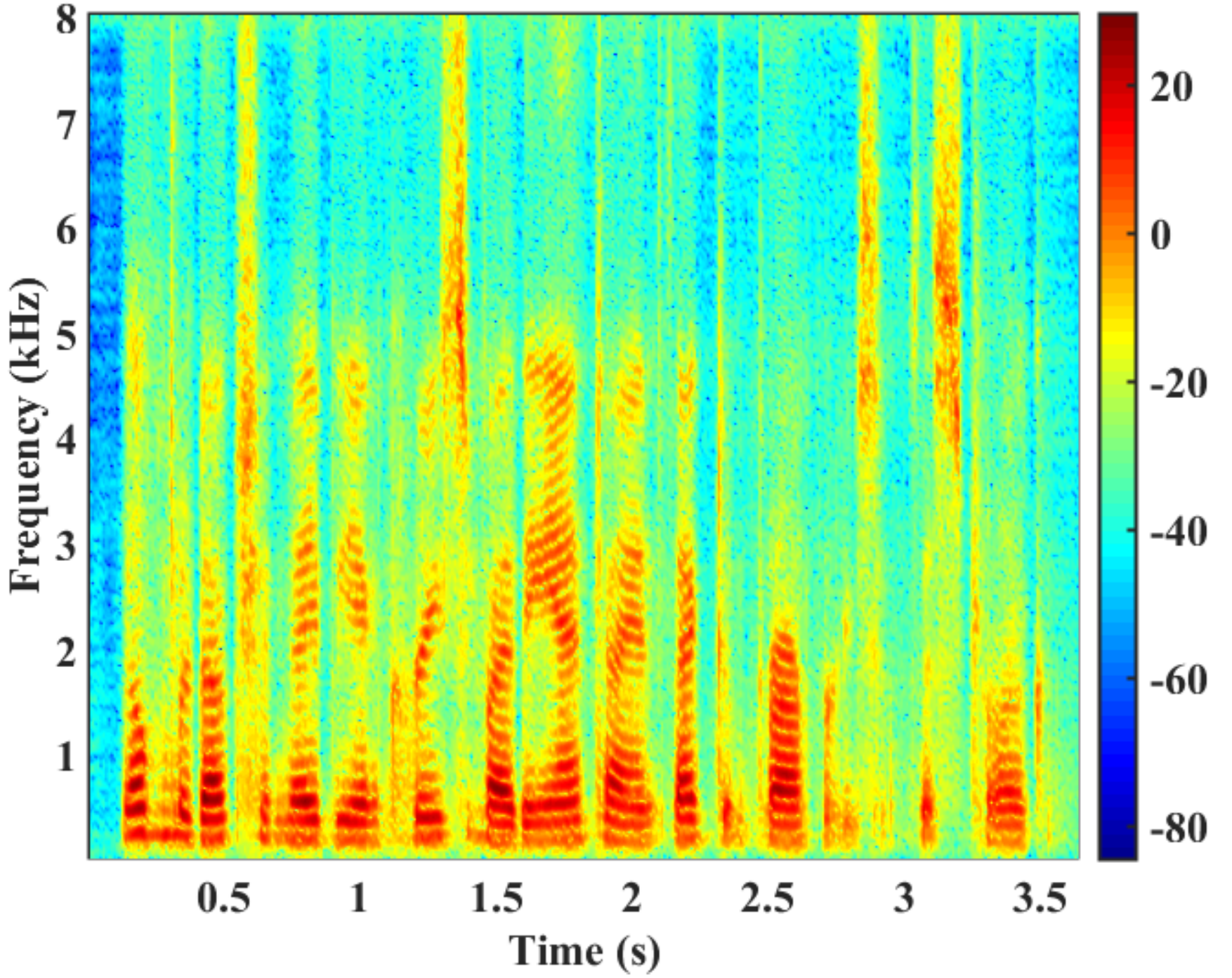}	
	}
	\caption{Speech spectrograms obtained using different models under 0{\rm dB} with the unseen factory2 noise. The utterance is selected from the testing dataset and TMS is the training target. (a) Noisy utterance, PESQ = 1.85, STOI = 0.76. (b) Clean utterance, PESQ = 4.50, STOI = 1.00. (c) PL-DNN, PESQ = 2.36, STOI = 0.87. (d) PL-LSTM, PESQ = 2.44, STOI = 0.88. (e) CRNN, PESQ = 2.55, STOI = 0.89. (f) PL-CRNN, PESQ = 2.60, STOI = 0.90.}
	\label{fig:spectrogram-visualization}
	\vspace{0cm}
\end{figure}

\subsubsection{Comparison with CRNNs}
In this section, we compare the performance between CRNNs and PL-CRNN with $Q = 5$. The following phenomena can be observed. (1) PL-CRNN obtains the overall superiority compared with CRNN but with much fewer parameters, which will be further analyzed in the next part. For example, going from CRNN to PL-CRNN, an average PESQ improvement is about 0.04 for unseen case, which proves the effectiveness of PL-CRNN with the increase of the number of stages. There are two main reasons. First, the use of progressive learning is helpful to compensate the performance gap in each stage where fewer channels and LSTM units are set. Second, the intermediate mapping results actually serve as the prior information to boost the subsequent mapping process. (2) SCRNN has the same architecture as CRNN except that the number of parameters is reduced to 25$\%$. One can get that SCRNN decreases PESQ by average 0.07 and STOI by 0.95\% compared with CRNN for seen noise. This results confirm that the performance of CRNN may reduce its performance with less parameters.

Speech spectrograms are presented in Figure~{\ref{fig:spectrogram-visualization}}. It is obvious that, compared with PL-DNN and PL-LSTM, both CRNN and PL-CRNN can better reconstruct the spectral details of the speech signal while both of them can effectively remove the non-stationary noise components.

\subsection{Model Comparison}
To compare the above mentioned models, we introduce the principle of calculation in terms of floating-point fused multiply-adds (FMAs) for different networks~{\cite{pruning2017}}. Note that FMAs of BN and nonlinear activation functions are few so that they are often neglected during evaluating the computational complexity.

As convolutional layer and fully-connected layer are two most common layers in neural network, their calculation formulas are given in the Table~{\ref{tbl:flops}}, where $T$ and $F$ denote the size of output feature map in time dimension and that in feature dimension, respectively. $C_{in}$ and $C_{out}$ denote the number of channels for the input and the output feature maps, respectively. $K$ refers to the kernel size. Note that $+1$ indicates that the bias is also considered. $F_{i}$ and $F_{o}$ are respectively the number of input neurons and that of output neurons. RNN unit is not considered separately as it can be essentially viewed as a type of FC operation with recurrent connection. Accordingly, FMAs can be given by:
\begin{equation}
\setlength{\abovedisplayskip}{2pt}
\setlength{\belowdisplayskip}{2pt}
\mathcal{O} = \sum_{i=1}^{\mathcal{I}} \{ \mathcal{O}\left( Conv_{i} \right), \mathcal{O}\left( FC_{i} \right) \},
\end{equation}
where $\mathcal{I}$ refers to the layer set.

\renewcommand\arraystretch{1.0}
\newcolumntype{V}{!{\vrule width 1pt}}
\begin{table*}[t]
	\caption{Calculation formulas of FMA for convolutional layer and fully-connected layer.}
	\centering
	\footnotesize
	
	\begin{tabular}{c|c}
		\hline\hline
		\textbf{Unit} & \textbf{Formula}\\
		\hline
		Convolutional layer (Conv) &$\mathcal{O}\left(TF\left( C_{in}K_{T}K_{F} + 1 \right)C_{out}\right)$\\
		\hline
		Fully-connected layer (FC) &$\mathcal{O}\left(F_{i}F_{o}\right)$\\
		\hline\hline
	\end{tabular}
	\label{tbl:flops}
	\vspace{0.1cm}
\end{table*}
	
The number of parameters and FMAs for different architectures are given in Table~{\ref{tbl:parameter-flops}}. As Table~{\ref{tbl:parameter-flops}} shows, compared with the baselines, PL-CRNN has the smallest number of trainable parameters, which indicates the highest parameter efficiency of PL-CRNN. In addition, We notice that PL-CRNN has slightly more FMAs than SCRNN when $Q=5$. This is because FMAs has a linear relationship with the number of stages $Q$, which can be dramatically decreased by reducing $Q$. However, note that PL-CRNN with $Q=3$ has the smaller FMAs than SCRNN, while its performance is much better. Besides we observe that a larger number of parameters with multiple-stacked LSTM layers do not bring obvious performance improvement, which explains the saturation effect of LSTM-only network~{\cite{gao2018densely, tan2018convolutional}}.

\renewcommand\arraystretch{0.7}
\newcolumntype{V}{!{\vrule width 1pt}}
\begin{table*}[t]
	\caption{The number of trainable parameters (NumOfParas) and FMAs. \textbf{Bold} indicates the smallest value is achieved for each case. The number of stages is $Q = 3$ for all PL-based models if no emphasis is given. The unit is million.}
	\centering
	\footnotesize
		
		\begin{tabular}{c|c|c}
			\hline\hline
			\textbf{Models} & \textbf{NumOfParas} & \textbf{FMAs} \\
			\hline
			\textbf{PL-DNN} &17.87 & 17.87\\
			\hline
			\textbf{PL-LSTM} &42.25 & 42.25\\
			\hline
			\textbf{CRNN} &17.59 & 25.28\\
			\hline
			\textbf{SCRNN} &4.40 & 6.34\\
            \hline
			\textbf{PL-CRNN} &$\mathbf{1.22}$ &$\mathbf{5.94}$  \\
			\hline
			\textbf{PL-CRNN (Q=5)} &1.33 &9.94  \\
			\hline\hline
	\end{tabular}
	\label{tbl:parameter-flops}
	\vspace{0.1cm}
\end{table*}

\section{Conclusions}
In this paper, we propose a progressive learning framework for CRNN, which takes advantage of both CNN and RNN to significantly reduce the number of parameters when compared with PL-DNN, PL-LSTM, and CRNN. Experimental results show that the proposed PL-CRNN algorithm obtains consistently better performance than the competing algorithms in terms of PESQ, STOI, and SDR.




\section*{References}


\begin{thebibliography}{99}
	\bibitem{loizou2013speech}
	P.~C. Loizou, Speech enhancement: theory and practice, CRC press, 2013.
	
	\bibitem{boll1979suppression}
	S.~Boll, Suppression of acoustic noise in speech using spectral subtraction,
	IEEE Transactions on acoustics, speech, and signal processing. 27(2) (1979)
	113--120.
	
	\bibitem{zheng13}
	H.~Hu, S.~Wang, C.~Zheng, X.~Li, A cepstrum-based preprocessing and postprocessing for speech enhancement in adverse environments, Applied Acoustics. 74(12) (2013) 1458--1462.
	
	\bibitem{chen2006new}
	J.~Chen, J.~Benesty, Y.~Huang, S.~Doclo, New insights into the noise reduction
	wiener filter, IEEE Transactions on audio, speech, and language processing.
	14(4) (2006) 1218--1234.
	
	\bibitem{zheng14}
	C.~Zheng, R.~Peng, X.~Li, A Constrained MMSE LP Residual Estimator for Speech Dereverberation in Noisy Environments, IEEE Signal Processing Letters. 21(12) (2014) 1462--1466.
	
	\bibitem{jensen1995reduction}
	S.~H. Jensen, P.~C. Hansen, S.~D. Hansen, J.~A. Sorensen, Reduction of
	broad-band noise in speech by truncated qsvd, IEEE Transactions on Speech and
	Audio Processing. 3(6) (1995) 439--448.
	
	\bibitem{healy2013algorithm}
	E.~W. Healy, S.~E. Yoho, Y.~Wang, D.~Wang, An algorithm to improve speech
	recognition in noise for hearing-impaired listeners, The Journal of the
	Acoustical Society of America. 134(4) (2013) 3029--3038.
	
	\bibitem{xu2014regression}
	Y.~Xu, J.~Du, L.-R. Dai, C.-H. Lee, A regression approach to speech enhancement
	based on deep neural networks, IEEE/ACM Transactions on Audio, Speech, and
	Language Processing. 23(1) (2014) 7--19.
	
	\bibitem{xu2014global}
	Y.~Xu, J.~Du, L.-R. Dai, C.-H. Lee, Global variance equalization for improving
	deep neural network based speech enhancement, in: 2014 IEEE China Summit \&
	International Conference on Signal and Information Processing (ChinaSIP),
	IEEE, 2014, pp. 71--75.
	
	\bibitem{fu2017complex}
	S.-W. Fu, T.-y. Hu, Y.~Tsao, X.~Lu, Complex spectrogram enhancement by
	convolutional neural network with multi-metrics learning, in: 2017 IEEE 27th
	International Workshop on Machine Learning for Signal Processing (MLSP),
	IEEE, 2017, pp. 1--6.
	
	\bibitem{wang2018supervised}
	D.~Wang, J.~Chen, Supervised speech separation based on deep learning: An
	overview, IEEE/ACM Transactions on Audio, Speech, and Language Processing.
	26(10) (2018) 1702--1726.
	
	\bibitem{williamson2016complex}
	D.~S. Williamson, Y.~Wang, D.~Wang, Complex ratio masking for monaural speech
	separation, IEEE/ACM Transactions on Audio, Speech and Language Processing
	(TASLP). 24(3) (2016) 483--492.
	
	\bibitem{wang2014training}
	Y.~Wang, A.~Narayanan, D.~Wang, On training targets for supervised speech
	separation, IEEE/ACM transactions on audio, speech, and language processing.
	22(12) (2014) 1849--1858.
	
	\bibitem{wang2005ideal}
	D.~Wang, On ideal binary mask as the computational goal of auditory scene
	analysis, in: Speech separation by humans and machines, Springer, 2005, pp.
	181--197.
	
	\bibitem{hummersone2014ideal}
	C.~Hummersone, T.~Stokes, T.~Brookes, On the ideal ratio mask as the goal of
	computational auditory scene analysis, in: Blind source separation, Springer,
	2014, pp. 349--368.
		
	\bibitem{park2017fully}
	S.~R. Park, J.~W. Lee, A fully convolutional neural network for speech
	enhancement, Proc. Interspeech 2017 (2017) 1993--1997.
	
	\bibitem{fu2016snr}
	S.-W. Fu, Y.~Tsao, X.~Lu, SNR-aware convolutional neural network modeling for
	speech enhancement., in: Interspeech, 2016, pp. 3768--3772.
	
	\bibitem{weninger2015speech}
	F.~Weninger, H.~Erdogan, S.~Watanabe, E.~Vincent, J.~Le~Roux, J.~R. Hershey,
	B.~Schuller, Speech enhancement with lstm recurrent neural networks and its
	application to noise-robust asr, in: International Conference on Latent
	Variable Analysis and Signal Separation, Springer, 2015, pp. 91--99.
	
	\bibitem{chen2017long}
	J.~Chen, D.~Wang, Long short-term memory for speaker generalization in
	supervised speech separation, The Journal of the Acoustical Society of
	America. 141(6) (2017) 4705--4714.
	
	\bibitem{zhao2018convolutional}
	H.~Zhao, S.~Zarar, I.~Tashev, C.-H. Lee, Convolutional-recurrent neural
	networks for speech enhancement, in: 2018 IEEE International Conference on
	Acoustics, Speech and Signal Processing (ICASSP), IEEE, 2018, pp. 2401--2405.
	
	\bibitem{gao2016snr}
	T.~Gao, J.~Du, L.-R. Dai, C.-H. Lee, SNR-based progressive learning of deep
	neural network for speech enhancement, in: INTERSPEECH, 2016, pp.
	3713--3717.
	
	\bibitem{gao2018densely}
	T.~Gao, J.~Du, L.-R. Dai, C.-H. Lee, Densely connected progressive learning for
	LSTM-based speech enhancement, in: 2018 IEEE International Conference on
	Acoustics, Speech and Signal Processing (ICASSP), IEEE, 2018, pp. 5054--5058.
	
	\bibitem{tan2018gated}
	K.~Tan, J.~Chen, D.~Wang, Gated residual networks with dilated convolutions for
	monaural speech enhancement, IEEE/ACM Transactions on Audio, Speech, and
	Language Processing. 27(1) (2018) 189--198.
	
	\bibitem{tan2018convolutional}
	K.~Tan, D.~Wang, A convolutional recurrent neural network for real-time speech
	enhancement., in: Interspeech, 2018, pp. 3229--3233.
	
	\bibitem{weninger2014discriminatively}
	F.~Weninger, J.~R. Hershey, J.~Le~Roux, B.~Schuller, Discriminatively trained
	recurrent neural networks for single-channel speech separation, in: 2014 IEEE
	Global Conference on Signal and Information Processing (GlobalSIP), IEEE,
	2014, pp. 577--581.
	
	\bibitem{vanwavenet}
	A.~van~den Oord, S.~Dieleman, H.~Zen, K.~Simonyan, O.~Vinyals, A.~Graves,
	N.~Kalchbrenner, A.~Senior, K.~Kavukcuoglu, Wavenet: A generative model for
	raw audio, in: 9th ISCA Speech Synthesis Workshop, pp. 125--125.
	
	\bibitem{ioffe2015batch}
	S.~Ioffe, C.~Szegedy, Batch normalization: Accelerating deep network training
	by reducing internal covariate shift, in: International Conference on Machine
	Learning, 2015, pp. 448--456.
	
	\bibitem{clevert2015fast}
	D.-A. Clevert, T.~Unterthiner, S.~Hochreiter, Fast and accurate deep network
	learning by exponential linear units (elus), arXiv preprint arXiv:1511.07289.
	
	\bibitem{noh2015learning}
	H.~Noh, S.~Hong, B.~Han, Learning deconvolution network for semantic
	segmentation, in: Proceedings of the IEEE international conference on
	computer vision, 2015, pp. 1520--1528.
	
	\bibitem{kim2016deeply}
	J.~Kim, J.~Kwon~Lee, K.~Mu~Lee, Deeply-recursive convolutional network for
	image super-resolution, in: Proceedings of the IEEE conference on computer
	vision and pattern recognition, 2016, pp. 1637--1645.
	
	\bibitem{ren2019progressive}
	D.~Ren, W.~Zuo, Q.~Hu, P.~Zhu, D.~Meng, Progressive image deraining networks: a
	better and simpler baseline, in: Proceedings of the IEEE Conference on
	Computer Vision and Pattern Recognition, 2019, pp. 3937--3946.
	
	\bibitem{zheng2015improving}
	H.~Zheng, Z.~Yang, W.~Liu, J.~Liang, Y.~Li, Improving deep neural networks
	using softplus units, in: 2015 International Joint Conference on Neural
	Networks (IJCNN), IEEE, 2015, pp. 1--4.
	
	\bibitem{hu2010tandem}
	G.~Hu, D.~Wang, A tandem algorithm for pitch estimation and voiced speech
	segregation, IEEE Transactions on Audio, Speech, and Language Processing.
	18(8) (2010) 2067--2079.
	
	\bibitem{varga1993assessment}
	A.~Varga, H.~J. Steeneken, Assessment for automatic speech recognition: Ii.
	noisex-92: A database and an experiment to study the effect of additive noise
	on speech recognition systems, Speech communication. 12(3) (1993) 247--251.
	
	\bibitem{kingma2014adam}
	D.~P. Kingma, J.~Ba, Adam: A method for stochastic optimization, arXiv preprint
	arXiv:1412.6980.
	
	\bibitem{rix2001perceptual}
	A.~W. Rix, J.~G. Beerends, M.~P. Hollier, A.~P. Hekstra, Perceptual evaluation
	of speech quality (pesq)-a new method for speech quality assessment of
	telephone networks and codecs, in: 2001 IEEE International Conference on
	Acoustics, Speech, and Signal Processing. Proceedings (Cat. No. 01CH37221),
	Vol.~2, IEEE, 2001, pp. 749--752.
	
	\bibitem{taal2011algorithm}
	C.~H. Taal, R.~C. Hendriks, R.~Heusdens, J.~Jensen, An algorithm for
	intelligibility prediction of time--frequency weighted noisy speech, IEEE
	Transactions on Audio, Speech, and Language Processing. 19(7) (2011)
	2125--2136.
	
	\bibitem{vincent2006performance}
	E.~Vincent, R.~Gribonval, C.~F{\'e}votte, Performance measurement in blind
	audio source separation, IEEE transactions on audio, speech, and language
	processing. 14(4) (2006) 1462--1469.
	
	\bibitem{pruning2017}
	H.~Li, A.~Kadav, I.~Durdanovic, H.~Samet, and H.~P.~Graf, Pruning filters for efficient convnets, in: ICLR, 2017.
	
\end{thebibliography}

\end{document}